\documentstyle[12pt]{article}

\begin{document}

\begin{titlepage}

{\tenrm\baselineskip=12pt
\hfill
 \parbox[t]{28mm}{SLAC--PUB--6000 \\
                 November 1992 \\
                 (T/E) }
}

\begin{flushright}
 DOE/ER/40322-186 \\
 U. of Md. PP \#92-087
\end{flushright}
\vspace{10mm}
\renewcommand{\thefootnote}{\fnsymbol{footnote}}
\begin{center}
 {\large\bf On the Self-Consistency of Scale-Setting
  Methods\footnote[1]{Work
  supported in part by Department of Energy contract DE--AC03--76SF00515
  (SLAC) and contract DE--FG05--87ER--40322 (Maryland).}
 } \\
 \vspace{10mm}
 Stanley J. BRODSKY \\
 {\it Stanford Linear Accelerator Center\\
 Stanford, California 94305 (U.S.A.)} \\
 \vspace{5mm}
 and \\
 \vspace{5mm}
 Hung Jung LU \\
 {\it Department of Physics, University of Maryland \\
 College Park, Maryland 20742 (U.S.A.)}
\end{center}

 \vspace{25mm}
 \begin{flushleft}
  {\bf Abstract}
 \end{flushleft}
 We discuss various self-consistency conditions for
 scale-setting methods. We show that the widely used
 Principle of Minimum Sensitivity (PMS) is disfavored
 since it does not satisfy these requirements.

\vspace{5mm}
\begin{center}
 Submitted to {\it Phys. Lett. B}
\end{center}

\end{titlepage}

Perturbative results in Quantum Chromodynamics (QCD)
suffer from the well-known scale ambiguity problem
\cite{FAC,PMS,BLM}.
That is, given a physical quantity
\begin{equation}
R_N = r_0 \alpha^p(\mu)
    + r_1(\mu) \alpha^{p+1}(\mu)
    + \dots
    + r_N(\mu) \alpha^{p+N}(\mu)
\end{equation}
expanded to $N$-th order in a coupling constant $\alpha(\mu)$,
the renormalization scale $\mu$ must be specified in order to obtain a
definite prediction. Although the infinite series
$R_\infty$ is renormalization scale independent, at
finite order the scale dependences from $\alpha(\mu)$
and $r_i(\mu)$ do not exactly cancel, leading hence
to a scale ambiguity.\footnote{
    Related to this subject is the scheme ambiguity problem.
    That is, we can choose $\alpha$ in any scheme in
    the expansion of $R_N$. However, as discussed in
    Ref. \cite{HLU}, the scheme ambiguity problem can be effectively
    reduced to the scale-ambiguity problem, in the sense
    that if we have a perfect scale-setting method, we can
    freely transform the coupling constant from one scheme
    to another.
}

Various scale-setting procedures have been
proposed in the literature:

\begin{enumerate}
 \item {\sl Fastest Apparent Convergence} (FAC)
 \cite{FAC} \\
 This method chooses the renormalization scale $\mu$
 that makes the next-to-leading order coefficient
 vanish:
 \begin{equation}
  r_1(\mu)=0 .
 \end{equation}

 \item{\sl Principle of Minimum Sensitivity} (PMS)
 \cite{PMS} \\
 This method chooses $\mu$ at the stationary point
 of $R$:
 \begin{equation}
  \frac{d R_N}{d \mu} = 0 .
 \end{equation}
 Beyond the next-to-leading order,
 PMS actually requires the optimization
 of scheme parameters in addition to the
 renormalization scale.

 \item{\sl Brodsky-Lepage-Mackenzie} (BLM)
 \cite{BLM} \\
 We shall consider it here to be the condition
 of a vanishing $N_f$ term in the next-to-leading
 order coefficient, where $N_f$ the number of light-quark
 flavors. That is, if
 \begin{equation}
  r_1(\mu) = r_{10}(\mu) + r_{11}(\mu) N_f,
 \end{equation}
 where $r_{10}(\mu)$ and $r_{11}(\mu)$ are $N_f$ independent,
 then BLM chooses the scale $\mu$ given by the condition
 \begin{equation}
  r_{11}(\mu)=0 .
 \end{equation}

This prescription ensures that, as in quantum electrodynamics,
vacuum polarization contributions
due to fermion pairs are associated with the coupling constant
$\alpha(\mu)$ rather than the coefficients.

\end{enumerate}

 Due to the absence of all-order results in QCD, it is difficult
to judge the performance of the various scale-setting
methods. However, there are a number of self-consistency
requirements that can shed some light into the reliability
of these methods.

\begin{enumerate}

 \item {\sl Existence and Uniqueness of} $\mu$. \\
 Clearly, it is desirable to have a scale-setting
 method that guarantees these two features.

 \item {\sl Reflexivity.} \\
 Given a coupling constant (or effective charge, see Ref. \cite{FAC})
 $\alpha(\mu)$ specified
 at a scale $\mu$, we can express it in terms of
 itself, but specified at another scale $\mu'$:
 \begin{equation}
  \alpha(\mu) = \alpha(\mu')
              - \frac{\beta_0}{4\pi} \log(\mu^2/\mu'^2)
                \alpha^2(\mu')
              + \cdots,
  \label{EqnReflex}
 \end{equation}
 where $\beta_0 = 11 - 2 N_f /3$ is the first
 coefficient of the QCD beta function.

 If a scale-setting prescription
 is self-consistent, it
 should choose the unique value $\mu'=\mu$ on the right-hand side.
 Notice that when $\mu'$ is chosen to be $\mu$,
 the above equation reduces to a trivial identity.
 This is a very basic requirement, since if
 $\alpha(\mu)$ is known (say, experimentally measured
 for a range of scale $\mu$),
 and then we try to use the above
 equation to ``predict" $\alpha(\mu)$ from itself, any deviation
 of $\mu'$ from $\mu$ would lead to an inaccurate result
 due to the truncation of the expansion series.

 \item {\sl Symmetry.} \\
 Given two different coupling constants
 (or effective charges \cite{FAC}, or renormalization schemes)
 $\alpha_1(\mu_1)$ and $\alpha_2(\mu_2)$, we can
 express one of them in terms of the other:
 \begin{eqnarray}
  \alpha_1(\mu_1) &=& \alpha_2(\mu_2)
                  + r_{12} \ (\mu_1,\mu_2) \alpha_2^2(\mu_2)
                  + \cdots,
                  \label{EqnSym1} \\
  \alpha_2(\mu_2) &=& \alpha_1(\mu_1)
                  + r_{21} \ (\mu_2,\mu_1) \alpha_1^2(\mu_1)
                  + \cdots.
                  \label{EqnSym2}
 \end{eqnarray}
 If a scale-setting method gives
 \begin{equation}
  \mu_2= \lambda_{21} \ \mu_1
 \end{equation}
 for the first series and
 \begin{equation}
  \mu_1= \lambda_{12} \ \mu_2
 \end{equation}
 for the second series, then this method is said to
 be symmetric if
 \begin{equation}
  \lambda_{12} \ \lambda_{21} = 1 .
  \label{SymCond}
 \end{equation}
 This feature is desirable since it gives us a unique
 ratio between $\mu_1$ and $\mu_2$, irrelevant of which way
 we choose to expand the coupling constants.

 \item {\sl Transitivity.} \\
 Given three different coupling constants
 $\alpha_1(\mu_1)$, $\alpha_2(\mu_2)$, and $\alpha_3(\mu_3)$,
 we can establish the relation between $\mu_1$ and $\mu_3$
 in two ways:
 \begin{enumerate}
  \item Going through the extra scheme $\alpha_2(\mu_2)$.
  That is, we can fix the scales in the two series
  \begin{eqnarray}
   \alpha_1(\mu_1) &=& \alpha_2(\mu_2)
                   + r_{12}(\mu_1,\mu_2) \ \alpha_2^2(\mu_2)
                   + \cdots, \\
   \alpha_2(\mu_2) &=& \alpha_3(\mu_3)
                   + r_{23}(\mu_2,\mu_3) \ \alpha_3^2(\mu_3)
                   + \cdots,
  \end{eqnarray}
  to obtain
  \begin{eqnarray}
   \mu_1 &=& \lambda_{12} \ \mu_2, \\
   \mu_2 &=& \lambda_{23} \ \mu_3,
  \end{eqnarray}
  and combine the last two expressions to get
  \begin{equation}
   \mu_1 = \lambda_{12} \lambda_{23} \ \mu_3 .
   \label{ExtraEvol}
  \end{equation}

  \item Directly setting $\mu_1$ in terms of $\mu_3$ in the
  series
  \begin{equation}
  \alpha_1(\mu_1)  = \alpha_3(\mu_3)
                   + r_{13}(\mu_1,\mu_3) \alpha_3^2(\mu_3)
                   + \cdots \\
  \end{equation}
  to get
  \begin{equation}
   \mu_1 = \lambda_{13} \ \mu_3 .
   \label{DirectEvol}
  \end{equation}
 \end{enumerate}
 A scale-setting method is transitive if
 Eqs. (\ref{ExtraEvol}) and (\ref{DirectEvol})
 give the same result. That is,
 \begin{equation}
  \lambda_{12} \ \lambda_{23} = \lambda_{13}.
  \label{TransCond}
 \end{equation}

\end{enumerate}

A scale-setting method that satisfies reflexivity, symmetry, and
transitivity effectively establishes an equivalent relation
among all the effective charges. This is a highly desirable
feature since it guarantees that no matter how we
move from one effective charge to another, we will always
keep a consistent choice of scale.

In what follows we will consider the case of QCD with
$N_f$ massless quarks.
It is straightforward to verify that the FAC and BLM criteria
satisfy all the consistency requirements outlined above.

\begin{enumerate}
 \item The existence and uniqueness of $\mu$ are guaranteed, since
 the scale-setting conditions
 of FAC and BLM are simple linear equations
 in $\log \mu^2$. In fact, the next-to-leading
 coefficient $r_1(\mu)$ in Eq. (1) has the form
 \begin{equation}
  r_1(\mu) = ( a + b N_f) + ( c + d N_f) \log \mu^2 ,
 \end{equation}
 with $a, b, c$ and $d$ simple constants that are independent
 of $N_f$. The solution given by FAC is
 \begin{equation}
  \log \mu_{\rm FAC}^2
  = - \frac{a+b N_f}{c+d N_f} ,
 \end{equation}
 whereas the solution given by BLM is
 \begin{equation}
  \log \mu_{\rm BLM}^2
  = - \frac{b}{d} .
 \end{equation}
 \item Reflexivity is satisfied. In Eq. (6) both methods
 require the logarithm $\log(\mu^2/\mu'^2)$
 in the next-to-leading coefficient to vanish, hence
 \begin{equation}
  \mu'=\mu .
 \end{equation}
 \item Symmetry is trivial since in Eqs.
 (\ref{EqnSym1}) and (\ref{EqnSym2})
 we always have
 \begin{equation}
  r_{12}(\mu_1,\mu_2) = -r_{21}(\mu_2,\mu_1).
 \end{equation}
 That is, the two next-to-leading order coefficients only
 differ by a sign. Thus, requiring one of them to vanish
 (FAC) is equivalent to requiring the other one to vanish,
 and requiring one of them to be $N_f$-independent is
 equivalent to requiring the other one to be $N_f$-independent.
 \item Transitivity also follows in both cases. In FAC the
 scales $\mu_1$ and $\mu_2$ are chosen such that the
 next-to-leading order term vanishes:
 \begin{eqnarray}
  \alpha_1(\mu_1) &=& \alpha_2(\mu_2) + O(\alpha_2^3),
  \label{EqFACTrans1} \\
  \alpha_2(\mu_2) &=& \alpha_3(\mu_3) + O(\alpha_3^3).
  \label{EqFACTrans2}
 \end{eqnarray}
 Substituting Eq.
 (\ref{EqFACTrans2}) into (\ref{EqFACTrans1}) we obtain
 \begin{equation}
  \alpha_1(\mu_1) = \alpha_3(\mu_3) + O(\alpha_3^3).
 \end{equation}
 Notice that this last equation does not contain the
 next-to-leading order term, either.
 Thus, the relationship between $\mu_1$ and $\mu_3$ is still given
 by the FAC condition (i.e., no next-to-leading order term),
 even when we have employed an intermediate
 scheme. For the BLM method we have a similar situation.
 If the scales $\mu_2$ and $\mu_3$ in the following two series
 \begin{eqnarray}
  \alpha_1(\mu_1) &=& \alpha_2(\mu_2)
                   + r_{12}(\mu_1,\mu_2) \alpha_2^2(\mu_2)
                   + O(\alpha_2^3),
  \label{Eq12} \\
  \alpha_2(\mu_2) &=& \alpha_3(\mu_3)
                   + r_{23}(\mu_2,\mu_3) \alpha_3^2(\mu_3)
                   + O(\alpha_3^3),
  \label{Eq23}
 \end{eqnarray}
 are chosen by the BLM method,
 then $r_{12}(\mu_1,\mu_2)$
 and $r_{23}(\mu_2,\mu_3)$ are independent of $N_f$. After
 replacing Eq. (\ref{Eq23}) into Eq. (\ref{Eq12})
 \begin{equation}
  \alpha_1(\mu_1)  = \alpha_3(\mu_3)
                   + [  r_{12}(\mu_1,\mu_2)
                      + r_{23}(\mu_2,\mu_3) ] \ \alpha_3^2(\mu_3)
                   + O(\alpha_3^3),
 \end{equation}
 we see that the next-to-leading order coefficient will
 also be $N_f$ independent, since it is the sum
 of two $N_f$-independent quantities.

\end{enumerate}

Now let us turn our attention to PMS. Unfortunately it does
not satisfy any of the self-consistency conditions outlined
previously.

To begin with, unlike the cases of FAC and BLM,
in general there are no known theorems that guarantee the existence
or the uniqueness of the PMS solution. Although for
practical cases PMS does provide solutions, and when
there are more than one solution usually only one
of them lies in the physically reasonable region \cite{PMS},
these observations alone do not prove that PMS will be
trouble-free for new processes.

Before analyzing the other self-consistency relations, let us
perform first some preliminary calculations.

Given the QCD beta function for an effective charge
$\alpha_1$:
\begin{equation}
\beta(\alpha_1) =
\frac{d}{d \log \mu_1^2}
\left( \frac{\alpha_1}{4\pi}
\right)
= -\beta_0
\left( \frac{\alpha_1}{4\pi}
\right)^2
-\beta_1
\left( \frac{\alpha_1}{4\pi}
\right)^3
- \cdots,
\end{equation}
where $\beta_0 = 11 - \frac{2}{3} N_f$ and
$\beta_1 = 102 - \frac{38}{3} N_f$.
To the next-to-leading order,
$\alpha_1$ is implicitly given by the
following equation:
\begin{equation}
\frac{1}{a_1}
+ \log \left( \frac{a_1}{1+a_1}
       \right)
=\tau_1
\end{equation}
where $a_1=\beta_1 \beta_0^{-1} \alpha_1/4\pi$,
and $\tau_1=\beta_0^2\beta_1^{-1}\log(\mu_1^2/\Lambda_1^2)$.
(The scale $\Lambda_1$ is effectively the 't Hooft scale
of $\alpha_1$. See Ref. \cite{LB}.)

Given two effective charges $\alpha_1$ and $\alpha_2$
(or, $a_1$ and $a_2$,)
they are related by the perturbative series
\begin{equation}
a_1(\tau_1) =
a_2(\tau_2)
+ (\tau_2-\tau_1)
a_2^2(\tau_2)
+ \cdots ,
\end{equation}
where $a_2 = \beta_1 \beta_0^{-1} \alpha_2/4\pi$,
and $\tau_2 = \beta_0^2 \beta_1^{-1}\log(\mu_2^2/\Lambda_2^2)$.
This is an equation of the form of Eq. (1) where the scale $\mu_2$
is to be chosen.
PMS proposes the choice of $\mu_2$ (or equivalently, $\tau_2$)
at the stationary point, {\it i.e.}:
\begin{equation}
\frac{d a_1}{d \tau_2}
= 0 =
\frac{d}{d \tau_2}
\left[ a_2(\tau_2)
      + (\tau_2-\tau_1)
        a_2^2(\tau_2)
\right] .
\end{equation}
{}From here we obtain the condition:
\begin{equation}
1+a_2
=
\frac{1}{2(\tau_1-\tau_2)} .
\end{equation}
In order to obtain $\tau_2$ in terms of $\tau_1$,
we must solve the last equation in conjunction with
\begin{equation}
\frac{1}{a_2}
+ \log \left( \frac{a_2}{1+a_2}
       \right)
=\tau_2 .
\end{equation}
In Fig. 1 we present the graphical solution of
the PMS scale-parameter $\tau_2$ as a function
of the external scale-parameter $\tau_1$.

\setlength{\unitlength}{1cm}
\begin{picture}(10,6)(-4.5,-1.0)

\put(0,0){\vector(1,0){4}}
\put(0,0){\vector(0,1){4}}

\put( 4.2,-0.3){$\tau_1$}
\put(-0.5, 4.0){$\tau_2$}
\put(-0.3,-0.3){$0$}
\put( 2.4, 0.4){$\tau_2=\tau_1-\frac{1}{2}$}
\put( 2.2, 0.5){\vector(-1,0){0.4}}

\put(0.2,0.0){\line(5,3){0.5}}
\put(0.7,0.3){\line(5,3){0.5}}
\put(1.2,0.6){\line(5,3){0.5}}
\put(1.7,0.9){\line(4,3){0.4}}
\put(2.1,1.2){\line(5,4){0.5}}
\put(2.6,1.6){\line(1,1){1.4}}

\put(1.0,0.0){\line(1,1){0.36}}
\put(1.5,0.5){\line(1,1){0.36}}
\put(2.0,1.0){\line(1,1){0.36}}
\put(2.5,1.5){\line(1,1){0.36}}

\end{picture}

{\tenrm\baselineskip=12pt
\hglue 2cm
 \parbox[t]{1cm}{Fig. 1}
 \
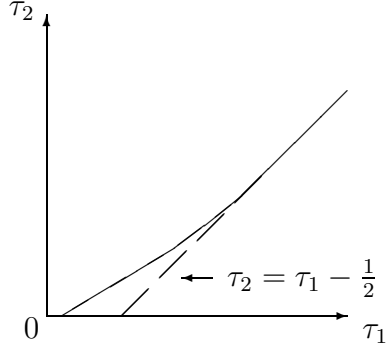
 \parbox[t]{8cm}{The dependence of the
 PMS scale parameter $\tau_2$ as a function
 the external scale parameter $\tau_1$.}

}
\vspace{0.5cm}

Notice that in the large momentum region $(\tau_1,\tau_2 \gg 1)$
we have
\begin{equation}
\tau_2 \sim
\tau_1 - \frac{1}{2} .
\end{equation}
In terms of $\mu_1$ and $\mu_2$, the relation becomes
\begin{equation}
\frac{\mu_2}{\Lambda_2}
\sim
\frac{\mu_1}{\Lambda_1}
\exp ( -\beta_1 /4 \beta_0^2 ) .
\label{PMSScales}
\end{equation}
Let us now check the other self-consistency relations
for PMS. For simplicity we will consider the large
momentum kinematic region where the
above approximation holds, although none of our conclusions
will rely on this approximation.

Reflexivity is violated in PMS. When the PMS method is
applied to Eq. (\ref{EqnReflex}), from
Eq. (\ref{PMSScales}) we obtain:
\begin{equation}
\mu' \sim \mu \exp(-\beta_1/4 \beta_0^2)
     \ne \mu .
\end{equation}
This is a severe drawback of the PMS method. When
we use an effective charge to predict itself,
the application of the PMS method would lead
to an inaccurate result. If PMS cannot provide
the optimum scale even in this simple situation,
its reliability for other processes is very
questionable.

Symmetry and transitivity are also violated in PMS.
{}From Eq. (\ref{PMSScales}) we know that in general:
\begin{equation}
\lambda_{ij} =
\frac{\mu_i}{\mu_j}
\sim
\frac{\Lambda_i}{\Lambda_j}
\exp(-\beta_1/4 \beta_0^2) .
\end{equation}
This would mean that
\begin{eqnarray}
  \lambda_{12} \lambda_{21}
  &\sim&
  \exp(-\beta_1/2 \beta_0^2)
  \ne 1 ,
  \\
  \lambda_{12} \lambda_{23}
  &\sim&
  \frac{\Lambda_1}{\Lambda_3}
  \exp(-\beta_1/2 \beta_0^2)
  \ne \lambda_{13}
  \sim
  \frac{\Lambda_1}{\Lambda_3}
  \exp(-\beta_1/4 \beta_0^2).
\end{eqnarray}
That is, PMS does not satisfy the symmetry and
transitivity conditions stated in Eqs.
(\ref{SymCond}) and (\ref{TransCond}).

Hence, when we successively express one effective
charge in terms of others, PMS would lead to
inconsistent scale choices. We can only conclude
that the PMS method in general does not
provide the optimum scale, since an optimum
scale-setting methods should satisfy all these
self-consistency requirements.

Let us point out that adding the
scheme-parameter optimization in PMS does
not change any of the above conclusions.
The inability of PMS to meet these self-consistency
requirements resides in that the derivative
operations in general do not commute with the
operations of reflexivity, symmetry and
transitivity.

Finally, let us briefly comment on the
extended renormalization group formalism
recently studied in Ref.
\cite{LB}. In this formalism, the effective charges of
two physical observables can be
related by an evolution path on the hypersurface
defined by the QCD universal coupling function
$a(\tau,\{ c_i \} )$
\cite{LB}, where $\tau$ is the scale parameter
and $\{ c_i \}$ are the scheme parameters.
Given a initial effective charge $a_{init}$ at the point
$(\tau_{init},\{ c_i^{init} \} )$,
we can use the evolution equations
\cite{PMS,LB}
\begin{eqnarray}
{\delta a \over \delta \tau}
&=& \beta(a,\{ c_i \} )
= -a^2 (1 + a + c_2 a^2 + c_3 a^3 + \cdots ) ,
\\
\frac{\delta a}{\delta c_n}
&=& \beta_{(n)} ( a, \{ c_i \} )
= -\beta   ( a, \{ c_i \} ) \int_0^a dx
   \frac{ x^{n+2} }
        {\beta^2 ( x, \{ c_i \} )} .
\end{eqnarray}
to evolve $a_{init}$ into a final effective charge
$a_{final}$ at the point $(\tau_{final},\{ c_i^{final} \} )$.\footnote{
   Due to the lack of knowledge of higher-order scheme
   parameter of the initial and final effective charges,
   in practice we need to truncate the fundamental beta function
   $\beta(a,\{ c_i \} )$ and solve $a(\tau,\{ c_i \} )$
   in a finite-dimensional subspace. }
As long as we stay inside an analytical region where
the second partial derivatives exist and commute, the
predicted value of $a_{final}$ will not depend on the
path chosen for the evolution.

In Fig. 2 we illustrate the paths that represent
the operations of reflexivity, symmetry and
transitivity. We can pictorially visualize that
the evolution paths satisfy all these three
self-consistency properties.
A closed path starting and
ending at the point $A$ represents the
operation of identity. Since the
predicted value does not depend on the chosen
path, if the effective charge at $A$ is $a_A$,
after completing the path we will also
end up with an effective charge $a_A$.
Similarly, if we evolve $a_B$ at $B$ to a
value $a_C$ at $C$, we are guaranteed that
when we evolve $a_C$ at $C$ back to the point $B$,
the result will be $a_B$. Hence, the evolution
equations also satisfy symmetry. Transitivity
follows in a similar manner. Going directly from
$D$ to $F$ gives the same result as
going from $D$ to $F$ through a third point $E$.

\setlength{\unitlength}{1cm}
\begin{picture}(10,6)(-6.0,-2.0)

\put(0,0){\vector(1,0){3}}
\put(0,0){\vector(0,1){3}}
\put(0,0){\vector(-2,-1){2}}

\put(-0.6, 3.2){$a(\tau,\{ c_i \} )$}
\put( 3.2,-0.3){$\tau$}
\put(-2.7,-1.4){$\{ c_i \} $}

\put(-0.4, 1.2){$A$}
\put(-0.5, 1.1){\circle*{0.1}}
\put(-0.7, 1.1){\circle{0.4}}
\put(-0.89, 1.19){\vector(0,1){0.0}}

\put( 0.4, 0.8){$B$}
\put(-1.2, 0.1){$C$}
\put( 0.6,0.6){\circle*{0.1}}
\put(-0.8,0.0){\circle*{0.1}}
\put( 0.6, 0.6){\line(-4,-3){0.4}}
\put( 0.2, 0.3){\vector(-2,-1){0.4}}
\put(-0.2, 0.1){\line(-6,-1){0.6}}

\put(-0.8, 0.0){\line( 4, 3){0.4}}
\put(-0.4, 0.3){\vector( 2, 1){0.4}}
\put( 0.0, 0.5){\line( 6, 1){0.6}}

\put( 1.3, 0.1){$D$}
\put( 1.1,-0.6){$E$}
\put(-0.4,-0.6){$F$}
\put( 1.2, 0.3){\circle*{0.1}}
\put( 0.9,-0.6){\circle*{0.1}}
\put( 0.0,-0.6){\circle*{0.1}}

\put( 1.2, 0.3){\vector(-1,-3){0.2}}
\put( 0.9,-0.6){\line( 1, 3){0.2}}
\put( 0.9,-0.6){\vector(-4, 0){0.4}}
\put( 0.0,-0.6){\line( 1, 0){0.5}}
\put( 1.2, 0.3){\vector(-4,-3){0.8}}
\put( 0.0,-0.6){\line( 4,3){0.4}}

\put(0.2,2.3){\line(2,-5){0.2}}
\put(0.4,1.8){\line(3,-5){0.3}}
\put(0.7,1.3){\line(1,-1){0.4}}
\put(1.1,0.9){\line(5,-3){0.5}}
\put(1.6,0.6){\line(2,-1){0.4}}
\put(2.0,0.4){\line(5,-2){0.5}}

\put(-1.8, 0.8){\line(2,-5){0.2}}
\put(-1.6, 0.3){\line(3,-5){0.3}}
\put(-1.3,-0.2){\line(1,-1){0.4}}
\put(-0.9,-0.6){\line(5,-3){0.5}}
\put(-0.4,-0.9){\line(2,-1){0.4}}
\put( 0.0,-1.1){\line(5,-2){0.5}}

\put(0.2,2.3){\line(-4,-3){2}}
\put(2.5,0.2){\line(-4,-3){2}}

\end{picture}

{\tenrm\baselineskip=12pt
\hglue 1cm
 \parbox[t]{1cm}{Fig. 2}
 \
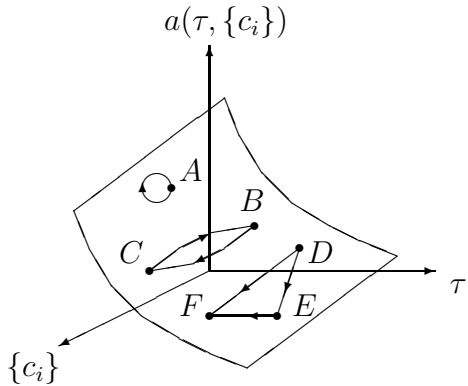
 \parbox[t]{9cm}{Pictorial
 representation of the universal coupling function.
 The point $A$ with a closed path
 represents the operation of reflexivity.
 The paths $\overline{BC}$ and $\overline{CB}$
 represent the operation of symmetry, and
 the paths $\overline{DE}$,$\overline{EF}$
 and $\overline{DF}$ represent the operation
 of transitivity.}

}
\vspace{0.5cm}
Summarizing, we have outlined a number of self-consistency
conditions for scale-setting methods, and shown
that FAC and BLM satisfy these requirements,
whereas PMS does not. We have pictorially argued
that the formalism based on the extended renormalization
group equations satisfies all these requirements
for scale and scheme variation.


\begin{thebibliography}{9}

\bibitem{FAC}
            G.~Grunberg,
            Phys.~Lett.~B    95,   70 (1980);
            Phys.~Lett.~B   110,  501 (1982);
            Phys.~Rev.~D     29, 2315 (1984).
\bibitem{PMS}
            P.~M.~Stevenson,
            Phys.~Lett.~B   100,   61 (1981);
            Phys.~Rev.~D     23, 2916 (1981);
            Nucl.~Phys.~B   203,  472 (1982);
            Nucl.~Phys.~B   231,   65 (1984).
\bibitem{BLM}
            S.~J.~Brodsky, G.~P.~Lepage and P.~B.~Mackenzie,
            Phys.~Rev.~D     28,  228 (1983).
\bibitem{HLU}
            H.~J.~Lu and C.~A.~R. S\'a de Melo,
            Phys.~Lett.~B   273,   260 (1991),
            erratum-ibid.B~285, 399 (1992).
\bibitem{LB} H.~J.~Lu and S.~J.~Brodsky,
             preprint DOE/ER/40322-178 (SLAC-PUB-5938),
             October, 1992.


\end{thebibliography}
\end{document}